\documentclass[]{aastex631}

\usepackage{xcolor}
\usepackage{colortbl}
\usepackage{color}
\graphicspath{{./}{figures/}}
\usepackage{hyperref}
\usepackage{float}
\usepackage{graphicx}
\usepackage{tablefootnote}

\begin{document}

\title{Multiple Emission Regions in Jets of Low Luminosity Active Galactic Nucleus in NGC 4278}

\author[0009-0005-6429-1921]{Samik Dutta}
\email{samikduttaphy@gmail.com}
\affiliation{University of Calcutta \\
92, Acharya Prafulla Chandra Road, Rajabazar, Kolkata, West Bengal 700009, India}
\affiliation{Birla Institute of Technology and Science – Pilani \\
Vidya Vihar, Pilani, Rajasthan 333031, India}
\author [0000-0002-1188-7503]{Nayantara Gupta}
\affiliation{Raman Research Institute \\
C. V. Raman Avenue, 5th Cross Road, Sadashivanagar, Bengaluru, Karnataka 560080, India}
\email{nayan@rri.res.in}

\begin{abstract}
The Large High Altitude Air Shower Observatory (LHAASO) has detected very high energy gamma rays from the LINER galaxy NGC 4278, which has a low luminosity active galactic nucleus, and symmetric mildly relativistic S-shaped twin jets detected by radio observations. Few low-luminosity active galactic nuclei are detected in gamma rays due to their faintness. Earlier, several radio-emitting components were detected in the jets of NGC 4278. We model their radio emission with synchrotron emission of ultra-relativistic electrons to estimate the strength of the magnetic field inside these components within a time-dependent framework after including the ages of the different components. We show that the synchrotron and synchrotron self-Compton emission by these components cannot explain the Swift X-ray data and the LHAASO gamma-ray data from NGC 4278. We suggest that a separate component in one of the jets is responsible for the high energy emission whose age, size, magnetic field and the spectrum of the ultra-relativistic electrons inside it have been estimated after fitting the multi-wavelength data of NGC 4278 with the sum of the spectral energy distributions from the radio components and the high energy component. We note that the radio components of NGC 4278 are larger than the high-energy component which has also been observed in several high-luminosity active galactic nuclei.
\end{abstract}


\section{Introduction} \label{sec:intro}
Low-ionisation nuclear emission-line regions (LINERs), first identified as a class of galaxies more than 40 years ago \citep{1980A&A....87..152H}, are the most abundant population of active galactic nuclei (AGNs) in our neighbourhood and the study of LINER galaxies could bridge our knowledge and understanding between normal and active galaxies \citep{10.3389/fspas.2017.00034}. The sources of this subclass of AGNs are identified by their low-ionisation, narrow emission lines from gas ionised by a non-stellar source \citep{Netzer_2015}. Due to their low accretion rate, they are faint and thus less often detected although they are abundant in nearby galaxies. 
Around 40\% of galaxies in the Palomar Spectroscopic Survey are AGNs, including LINER galaxies, Seyfert nuclei and transition objects \citep{Ho_1997b}.  H$\alpha$ components are detected in a large number of LINERs, in many cases points like X-ray or UV sources are also detectable at their nuclei, these LINERs are named LINER 1.9 \citep{Ho_1997c} and they form the sub-class of low luminosity AGNs (LLAGNs). Their average bolometric luminosity is less than $10^{42}$ erg/sec \citep{Terashima_2000}. It is also important to note that the commonly observed big blue bump from the standard accretion disc in more luminous AGNs is either absent or weak in LLAGNs \citep{2008ARA&A..46..475H}. 

\par
Earlier many LLAGNs have been described by advection-dominated accretion flows (ADAF), which are radiatively inefficient for sub-Eddington accretion rates with low densities and low optical depths. This results in a geometrically thick and optically thin accretion flow unlike geometrically thin and optically thick accretion flows in luminous AGNs \citep{1994ApJ...428L..13N}. Optically thin ADAF is usually referred to as radiatively inefficient accretion flow (RIAF).
The emission in LLAGNs may come from the jet, the RIAF and the outer thin disk \citep{10.1093/mnras/stt2388}. Their relative contribution may vary from one source to another. \cite{10.1111/j.1365-2966.2010.17224.x} modelled 24 LINERs and showed that both the ADAF model and jet model can explain the observed X-ray data.

\par
It is not known how similar are the LLAGNs to the more luminous AGNs. GeV energy gamma-rays have been detected by Fermi-LAT from NGC 315, NGC 4261, NGC 1275 and NGC 4486 \citep{2020MNRAS.492.4120D}. Multi-wavelength data analysis and modelling suggest that single zone synchrotron self-Compton (SSC) emission from the jet can explain the gamma-ray emission up to a few GeV from NGC 315 and NGC 4261 while hadronic emission from RIAF cannot explain the gamma-ray data. 
 These sources were further studied \citep{2021ApJ...919..137T} and the authors reported that SSC emission from sub-parsec scale jet can explain the gamma-ray data from all these four LLAGNs, however external Compton emission from the kilo-parsec scale jets of NGC 315 and NGC 4261 is needed to explain the gamma-ray data beyond 1.6 GeV and 0.6 GeV, respectively. In a more recent paper \citep{2023ApJ...950..113T} a detailed multi-wavelength study of the nearby LLAGN M 81* has been carried out at different quiescent and flaring states. The radio and X-ray data of different epochs are well fitted by synchrotron emission of ultra-relativistic electrons and the authors found this source has a similarity with high synchrotron peaked blazars. A comparison of the characteristics of the LLAGN NGC 4278 to BL Lac objects, Seyfert galaxies and LINERs has been discussed in the paper by \cite{Lian:2024xnb}.

\par
NGC 4278 is an elliptical galaxy located at a distance of 16.4 Mpc \citep{Tonry_2001}. \cite{Ho_1997c} classified this source as a LINER 1.9 by confirming the presence of broad H$\alpha$ line with a energy flux of $\log F(H\alpha)=-13.07$ erg cm$^{-2}$ sec$^{-1}$, hence it is a LLAGN. The mass of its black hole is 3.09 $\pm 0.58 \times 10^8 M_{\sun}$ (\cite{2003MNRAS.340..793W}; \cite{Chiaberge_2005}), this implies an Eddington luminosity of 3.9 $\times 10^{46}$ erg/sec. \cite{Younes_2010} carried out simultaneous and quasi-simultaneous X-ray and multi-wavelength study of this source and pointed out that at a low X-ray flux, the spectral energy distribution (SED) of NGC 4278 is similar to that of a typical LINER source where the radio to X-ray emission can be considered as originating from a jet and/or RIAF. In contrast, at a high state of X-ray emission, it is more like a low luminosity Seyfert. Hence, they concluded that NGC 4278 may appear as a LINER or a Seyfert depending on the intensity of its X-ray emission.

\par

\cite{Giroletti_2005} reported a detailed study of the radio-emitting components in the jets of NGC 4278 using the observational data from Very Long Baseline Array (VLBA) at 5 GHz and 8.4 GHz. Their analysis revealed a two-sided structure, with symmetric S-shaped jets emerging from a flat-spectrum core. Assuming that the radio morphology is intrinsically symmetric and Doppler beaming effects govern its appearance, they found that NGC 4278 has mildly relativistic jets ($\beta \sim 0.75$), one of them is closely aligned to the line of sight ($2^{\circ}\le \theta \le 4^{\circ}$). The alternative scenario suggested in their paper is the source could be oriented at a larger angle and asymmetries could be related to the jet interaction with the surrounding medium. The details of the radio observations from their paper are discussed in the next section. We have used the radio data recorded by VLBA in 2000 from the four components S1, S2, N3 and N2 in the jets of NGC 4278 and fitted their radio flux at 5 GHz and 8.4 GHz with synchrotron emission of ultra-relativistic electrons to estimate the magnetic fields inside these components within a time-dependent framework, where the electron spectra evolved with time. \cite{Giroletti_2005} provides the Doppler factor of the jet frame, the sizes of the radio components, and their ages in the year 2000, we have used them to determine their multi-wavelength SEDs with the SSC model.

\par
NGC 4278 has also been observed in X-ray by Swift-XRT and in gamma-ray energy bands by \href{http://agile.rm.iasf.cnr.it/}{AGILE} and \href{http://www-glast.stanford.edu/}{Fermi-LAT} (\cite{Baldini_2021}, \cite{2024ApJS..271...10W}).
 \href{http://english.ihep.cas.cn/lhaaso/}{ LHAASO} (Large High Altitude Air Shower Observatory) has also reported very high energy gamma-ray detection from a source 1LHAASO J1219+2915, which is located at a distance of 0.05$^{\circ}$ from NGC 4278 \citep{Cao_2024a}. More recently they have reported that the TeV source is located within 0.03 $^{\circ}$ of NGC 4278 \citep{LHAASO:2024qzv}. The observations by LHAASO - WCDA during the active period have a significance of 8.8 $\sigma$. In AGN studies identifying the region of emission in different frequencies is crucial to understanding the underlying mechanisms of jet emission. Sometimes correlation is found between emissions in different frequencies which helps us to unveil the populations of electrons responsible for the emissions and model the sources in detail. The low luminosity AGN population is more unexplored than the high energy one. It is important to know which component in the jets of NGC 4278 could be responsible for the high energy emission to understand the characteristics of this class of AGN. We have modelled the multi-wavelength data of NGC 4278 to identify a possible emission region of the X-rays and gamma-rays in one of the mildly relativistic jets.

\section{Radio Observation of Components in Jets of NGC 4278} \label{sec:radio}
NGC 4278 was observed with an 11-element VLBI array made up of the NRAO Very Long Baseline Array (VLBA) on August 27, 2000. The observations were conducted by switching between 5 GHz and 8.4 GHz. The initial calibration was performed by the NRAO Astronomical Image Processing System (AIPS) and similar data reduction techniques were followed for both 5 GHz and 8.4 GHz data sets. 
\par 
\cite{Giroletti_2005} reveals that NGC 4278 is dominated by a compact component C ($T_B = 1.5 \times 10^9 K$) at the centre with jets on either side of it. In the southeast, the jet extends for $\sim 6.5$ mas before turning east. This jet-like feature is $20$ mas $\sim 1.4$ pc long. On the other side, the jet-like feature is elongated to the north in the 5 GHz map and the 8.4 GHz map also clearly shows a secondary component. Finally, this northern jet-like feature turns into a diffuse, low-brightness region as it bends to the west. NGC 4278 extends over $\sim 45$ mas which corresponds to about $3$ pc. It is classified as a compact symmetric object (CSO), except that it is under-luminous when compared to most known CSOs \citep{2000ApJ...534...90P}. A total flux density of $120$ mJy at $5$ GHz and $95$ mJy at $8.4$ GHz is measured. This visibility data is fitted by a five-component model at both of these frequencies. The sizes of the jet emission regions and their flux densities at $5$ GHz \& $8.4$ GHz are calculated from the parameters of the Gaussian component for the model brightness distribution of the fitting. 
\par 
The jet components S1, S2, N2 and N3 are named based on the most likely epoch of ejection. Information on the evolution of the source is obtained by comparing data taken at the same frequency in different epochs. Reprocessed data from July 1995 ($5$ GHz) is overlayed with the data from August 2000 ($5$ GHz) to see any change in the source structure. The overall morphology of NGC 4278 is seen to be almost identical to \cite{Giovannini_2001}.
\par
For each component, the radial distance increased over the 5 years between observations with larger motions on the northwestern side. N2 was displaced the most, with the component moving at $\sim 0.17c$. This observation argues that the northern jet is the one pointing towards us. \cite{Giroletti_2005} found that this jet is oriented close to the line of sight ($2^{\circ} \lesssim \theta \lesssim 4^{\circ}$) and it has a mildly-relativistic velocity of $\beta \sim 0.76$ with a corresponding Doppler factor $\delta \sim 2.7$.
\par
The ages of the components have been derived after assuming a constant apparent velocity for each component of NGC 4278. S2 and N2 are reported to be about 29 and 25 years old with S1 being the oldest component (65.8 years old) and N3 being the youngest (8.3 years old) in 2000. The ages of the respective components, their sizes and the Doppler factor $\delta \sim 2.7$ (for the jet pointing towards us) of the jet frame have been used by us to model their multi-wavelength spectral energy distributions (SEDs). We have denoted them as the radio components in Fig \ref{fig:radio_comp}, Fig \ref{fig:time_n3}, Fig \ref{fig:add_sed} and Fig \ref{fig:total_sed_power}. The errors in the radio fluxes are not given in Table 2 of \cite{Giroletti_2005}. In Table 1 the error in the total flux is about 5\%, based on this information we have assumed the error in the flux of individual components is 5\%, which is not noticeable in our figures.

\subsection{Modeling the Radio Data from Individual Components} 
The size and age of the individual components S1, S2, N3, N2 are given in \cite{Giroletti_2005} which we have used to calculate the radio spectrum from each of them. We have assumed that electrons are accelerated inside these components to ultra-relativistic energies, and they lose energy due to synchrotron and SSC emission, as a result of this the radio spectrum is produced.
\begin{figure}[ht]
\gridline{\fig{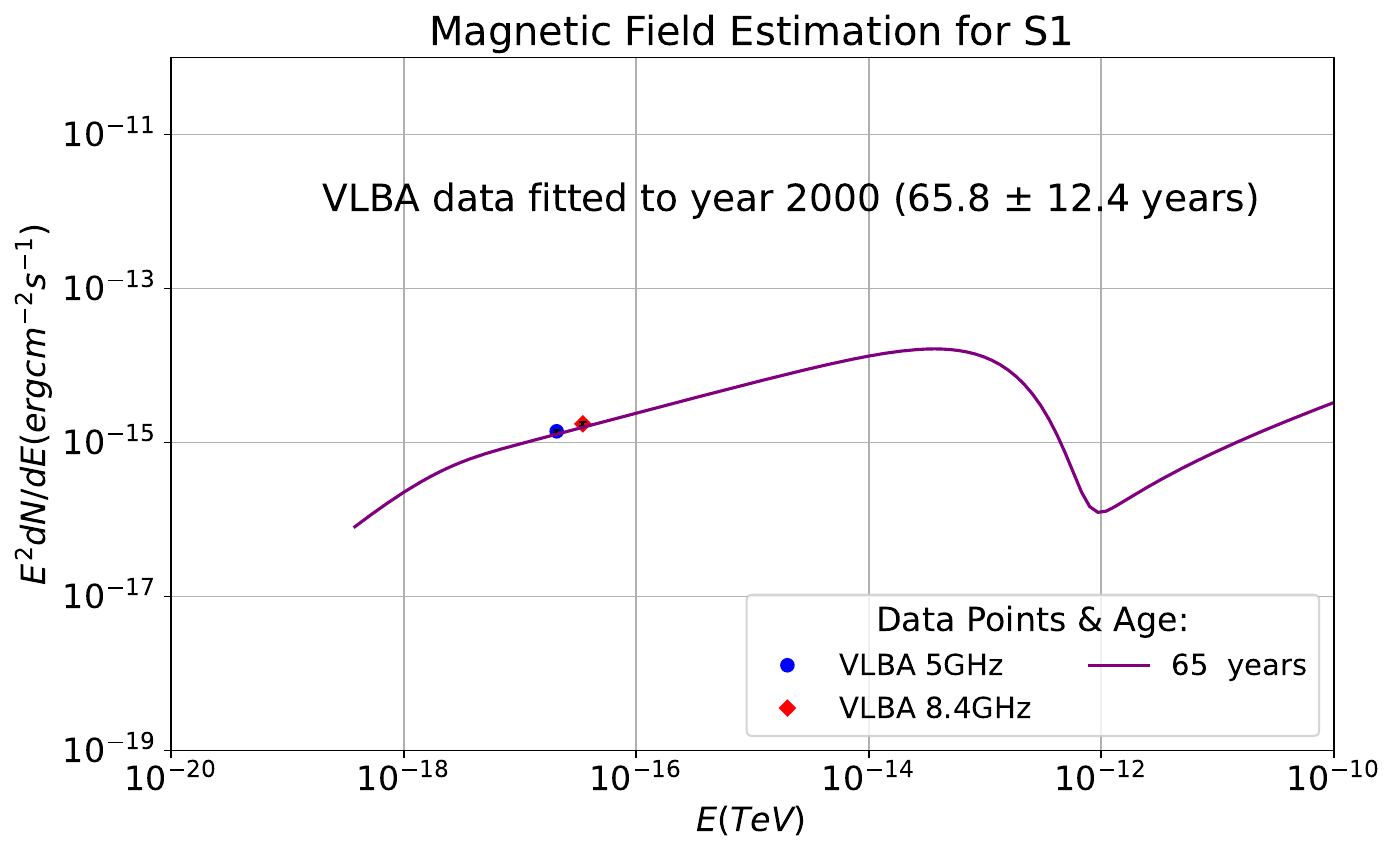}{0.5\linewidth}{(a)}\\
           \fig{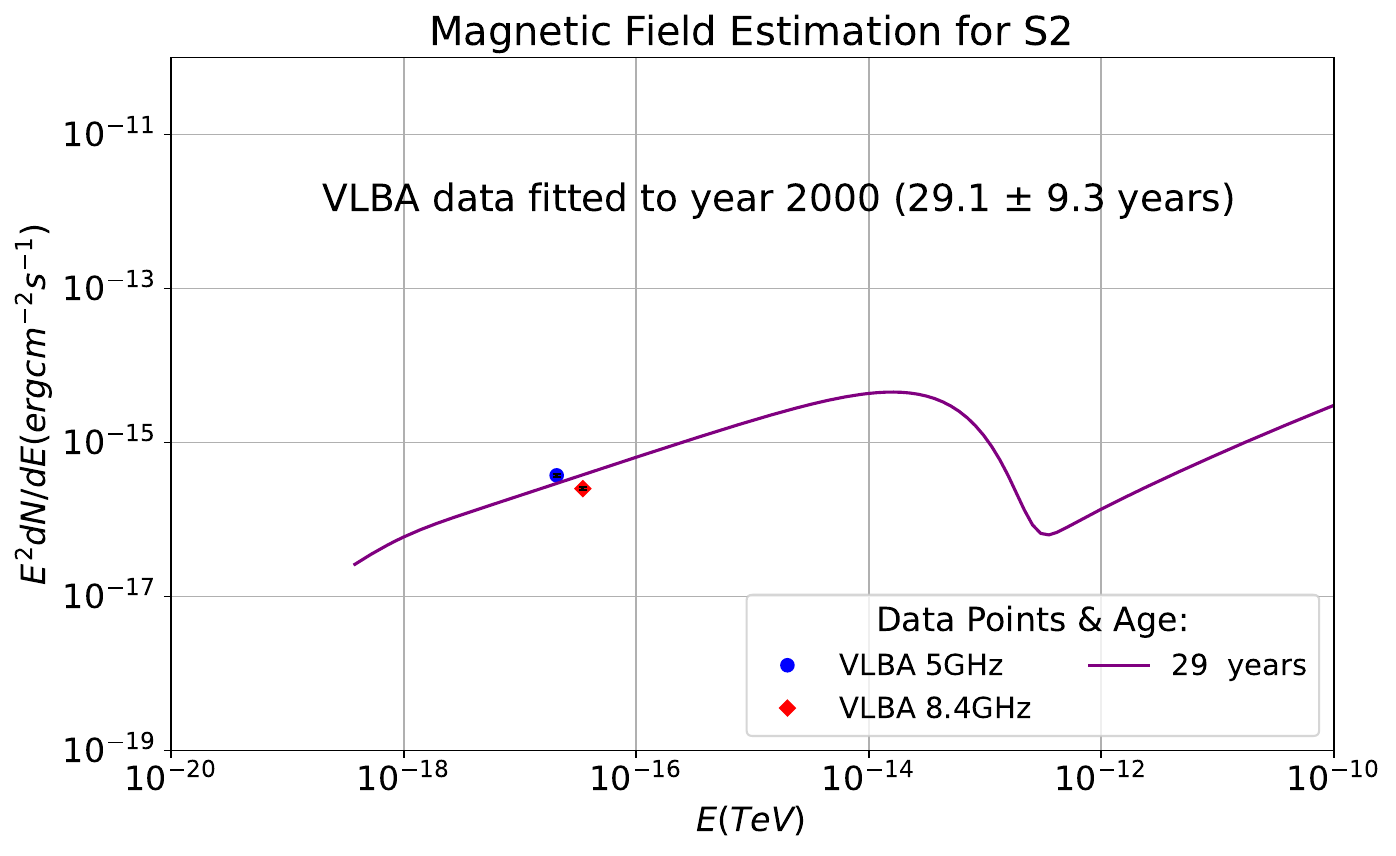}{0.5\linewidth}{(b)}\\}          
\gridline{\fig{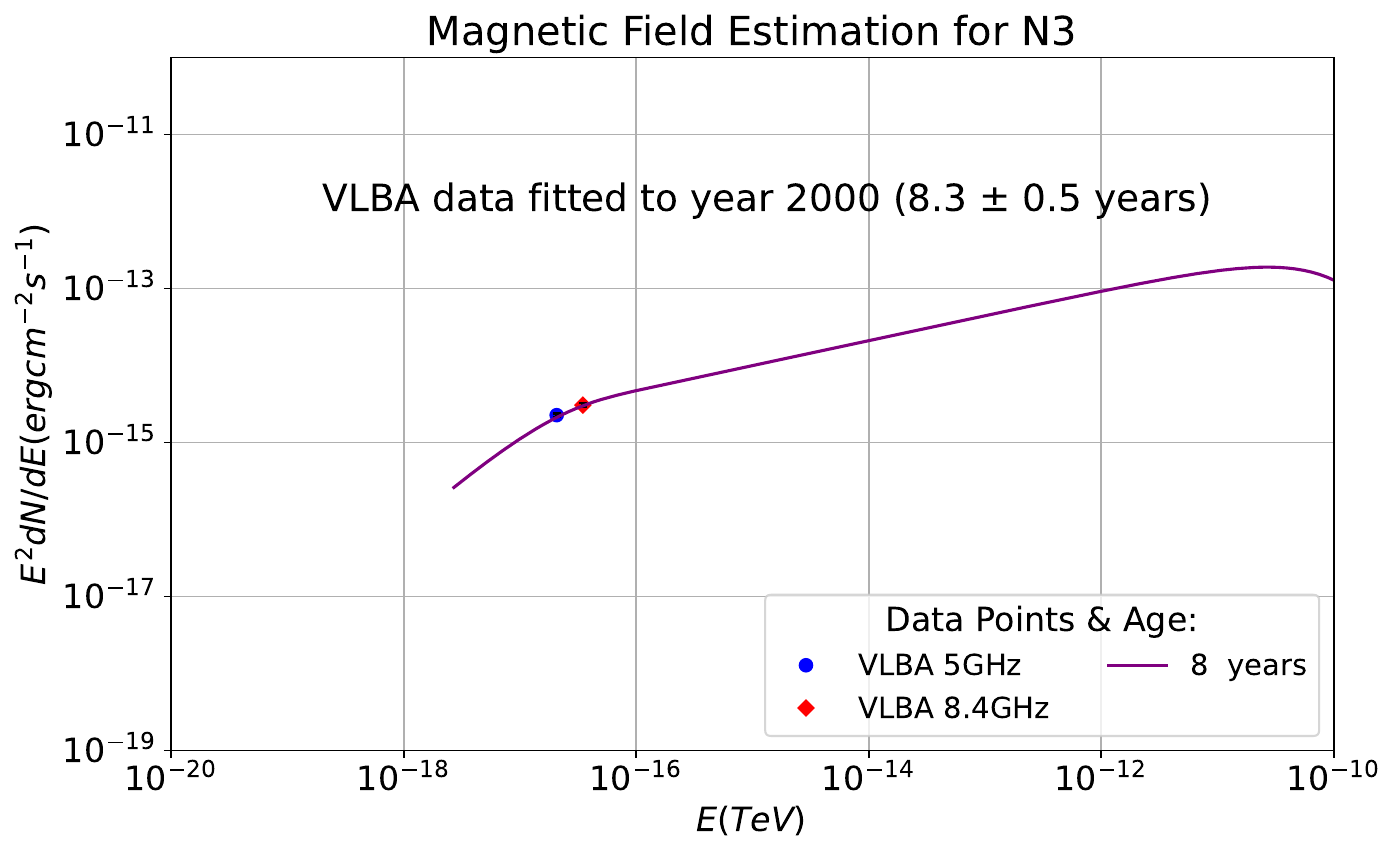}{0.5\textwidth}{(c)}\\
          \fig{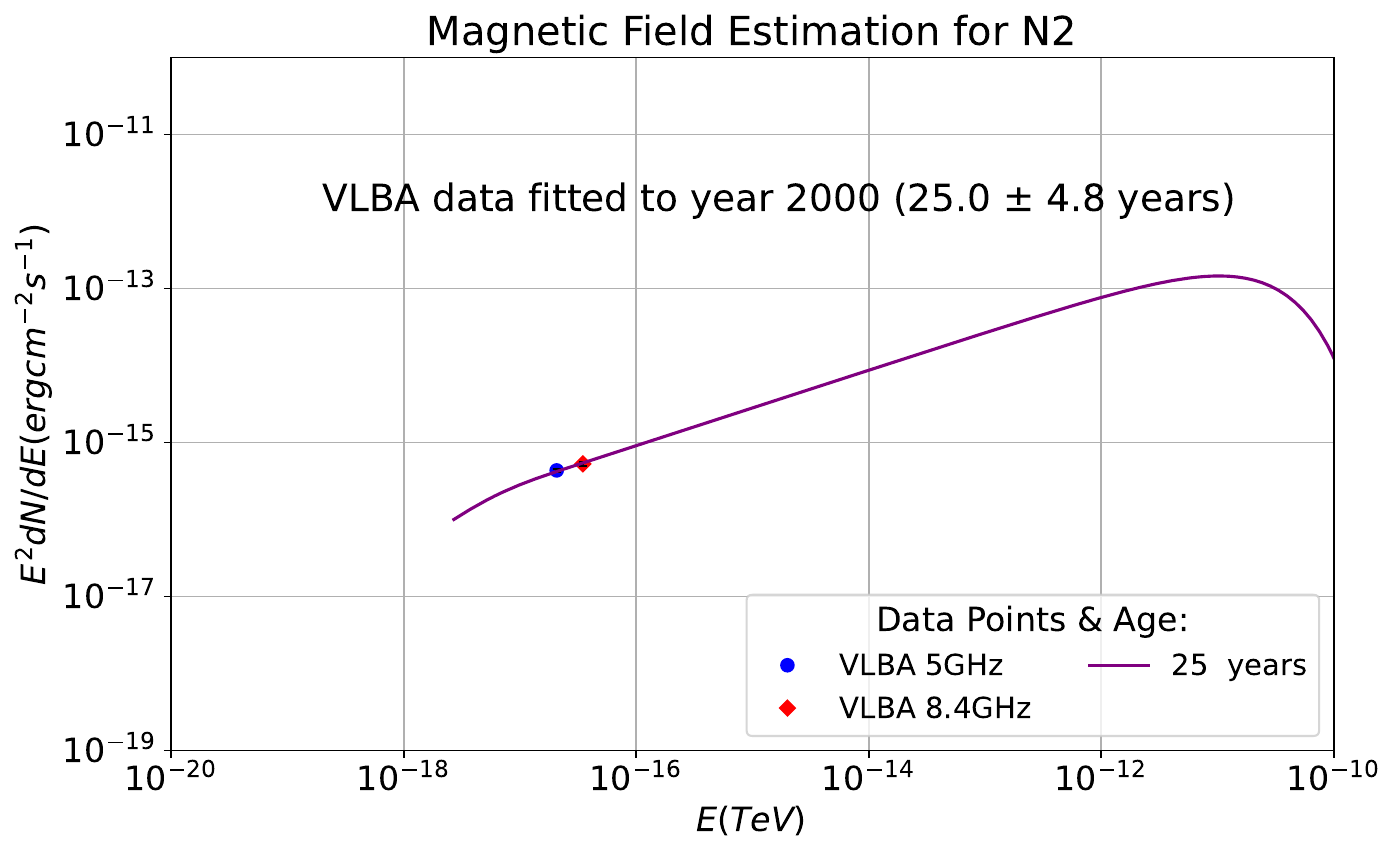}{0.5\textwidth}{(d)}\\}
\caption{Radio data of the year 2000 from the radio components reported in \cite{Giroletti_2005} have been fitted with synchrotron emission considering the time evolution of the accelerated electron spectra. The ages of the components in the year 2000 are also mentioned. The values of the parameters used to get the radio spectra are given in Table \ref{tab:parameter} of this paper. Errors on the radio data points are assumed to be 5\% following Table 1 of \cite{Giroletti_2005}.}
\label{fig:radio_comp}
\end{figure}

\begin{figure}[ht]
    \centering
    \includegraphics[width=0.9\linewidth]{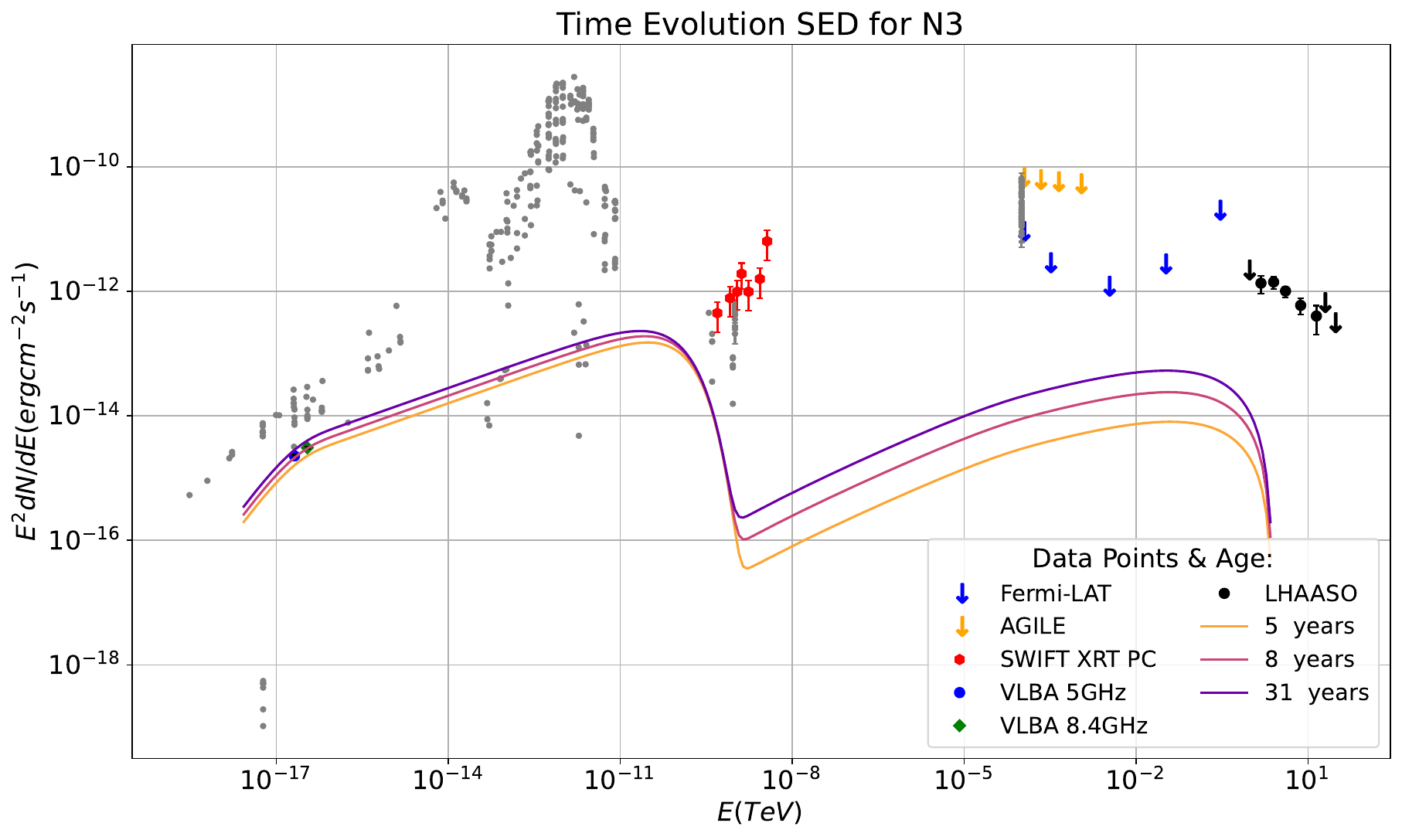}  
    \caption{The time evolution of the SED of the youngest radio component N3 shown in the years 2000 and 2023. Swift X-ray data (red solid circles) and Fermi-LAT upper limits (blue downward arrows) \citep{2024ApJS..271...10W}. The black downward arrows and black solid circles represent LHAASO upper limits and data points of NGC 4278 \citep{LHAASO:2024qzv}. The orange downward arrows represent the AGILE upper limits \citep{refId0}. The archival data points (grey solid circles) are from Energetic Gamma Ray Experiment Telescope (EGRET) \citep{1999ApJS..123...79H}, ROSAT All Sky Survey (RASS) \citep{1999A&A...349..389V}, WGACAT \citep{2000yCat.9031....0W} and NASA/IPAC Extragalactic Database (NED) catalogues.}
    \label{fig:time_n3}
\end{figure}

The GAMERA code \citep{2022ascl.soft03007H} has been used to solve the transport equation to generate the time-evolved electron spectrum inside each component. This code also calculates the resulting radiation spectra.
The transport equation for electrons is given by
\begin{equation}
\frac{\partial{N}}{\partial{t}}=Q(E,t)-\frac{\partial{(bN)}}{\partial{E}}-\frac{N(E,t)}{t_{esc}}
\label{eq:trnsp}
\end{equation}
 where $Q(E,t)$ is the injection spectrum of electrons produced by shock acceleration in an inhomogeneous magnetic field. 
A shock-accelerated electron spectrum may follow a simple power law distribution if the electrons are accelerated by the first-order Fermi mechanism. 
\begin{equation}
Q (E/E_0)= A \left(\frac{E}{E_0}\right)^{-\alpha}.
\label{eq:pw}
\end{equation}
The normalisation constant A is determined by the injected luminosity in electrons, $E_0$ is the reference energy, and $\alpha$ is the spectral index. We only have two data points for each radio component, which cannot constrain the spectral shape, however, the shape of the injected spectrum inside the radio components does not affect the result and conclusion of this work. Synchrotron and SSC are the most important energy loss processes in this case. These energy losses are included in the transport equation given in Eq.(\ref{eq:trnsp}) by the term $b(E,t)$. $t_{esc}$ denotes the timescale at which the electrons escape from the emission region. $N(E,t)$ is the resultant electron spectrum at any time t. The Doppler factor of the jet frame has been used from \cite{Giroletti_2005}.
In the first-order Fermi mechanism, the spectral index of the shock-accelerated electrons is close to -2, which has been used for the radio components. We have varied the magnetic field, and the normalisation of the electron spectrum to fit the radio data recorded in the year 2000 by VLBA at 5 GHz and 8.4 GHz from each of these components S1, S2, N3 and N2. The radio spectra and data points are shown in Fig \ref{fig:radio_comp}. We could not fit the radio data points of the S2 component as the flux measured at 8.4 GHz is lower than that at 5 GHz.
Subsequently, the electron spectra inside the components evolved in time till the year 2023 when all the values of the model parameters remained constant except the ages of the radio components. We have shown the time evolution of the multi-wavelength SED from the youngest component N3 in Fig \ref{fig:time_n3}. The multi-wavelength SED has not reached a steady state, hence a time-dependent model is needed for NGC 4278. In the years 2000 and 2023, this component was 8 years and 31 years old respectively. The SEDs are extended from radio to gamma-ray energy using the SSC model and added to get the total SED as shown in Fig. \ref{fig:add_sed}. The values of the parameters used to obtain the SEDs shown in Fig. \ref{fig:radio_comp}, Fig. \ref{fig:time_n3}  and  Fig. \ref{fig:add_sed} are given in Table \ref{tab:parameter}. The data from the year 2000 constrain the values of the parameters of the radio components. The archival radio data points are mostly above all components' total radio flux. 

\section{Modeling X-ray and Gamma-Ray Data from NGC 4278}
 The Swift X-ray and Fermi-LAT data were analysed by \cite{2024ApJS..271...10W}. They obtained some upper limits after reducing the Fermi-LAT data. We have used the Fermi-LAT upper limits and the Swift data points from their paper to compare with our model prediction. LHAASO is a multi-purpose extensive airshower array. It consists of three interconnected detector arrays, a km$^2$ array (KM2A) for gamma-ray detection above 10 TeV, a 78000 $m^2$ Water Cherenkov Detector Array (WCDA) for TeV gamma-ray detection and a Wide Field-of-View Cherenkov Telescope Array (WFCTA) for cosmic ray detection. Very high energy gamma-ray flux has been recorded using the WCDA by \cite{LHAASO:2024qzv}.
The  Swift X-ray \citep{2024ApJS..271...10W} and LHAASO gamma-ray data recorded by the WCDA \citep{LHAASO:2024qzv} cannot be fitted after adding the SEDs of the radio components S1, S2, N3 and N2 as shown in Fig \ref{fig:add_sed}.  We suggest that there is an additional component in the jet which is responsible for the high energy emission. This high-energy component has a smaller size and electrons are accelerated to very high energy within it. A power-law injection spectrum of ultra-relativistic electrons in the form of Eq. \ref{eq:pw} has been adapted to obtain the high energy photon emission radiated by the electrons. We solve the transport equation Eq. \ref{eq:trnsp} to get the time-evolved electron spectrum after including injection, energy loss and escape of electrons and subsequently the SED is obtained at the current age of the high energy component. The SED of synchrotron and SSC emission is shown in Fig \ref{fig:he_power}. After adding the SED of the high energy component to the SEDs of the radio components the total SED is shown in Fig \ref{fig:total_sed_power} and compared with the observational data. The values of the parameters used to fit the high energy data are listed in Table \ref{tab:parameterhigh}.

\section{Results}
We have fitted the VLBA radio data at 5 GHz and 8.4 GHz of the four components S1, S2, N3 and N2 in Fig \ref{fig:radio_comp}. 
The simulated spectra fit the radio data points of the radio components S1, N2 and N3, only for the 
 S2 component it is not possible to fit the radio data points. Although we cannot constrain the shape of the radio spectrum it does not affect the conclusion of our paper. We have used a simple power law spectrum of injected electrons in the radio components and showed that after time evolution the total multi-wavelength SED of the radio components cannot explain the X-ray and the gamma-ray data. Even if we assume a Log Parabola distribution for the injected electrons in the radio components
  the time-evolved electron spectra cannot explain the high energy radiation in X-ray and gamma-ray frequencies.
The size and the age of each component and the Doppler factor of the jet frame are used from \cite{Giroletti_2005}. The apparent velocities of the components are negligible compared to the jet velocity, so we have used the Doppler factor of the jet frame to calculate the SEDs from the components. The Doppler factor depends on the jet orientation angle ($\theta$), which is between 2$^{\circ}$ and 4$^{\circ}$ for the jet pointing towards us \citep{Giroletti_2005}, and about 177$^{\circ}$ for the jet pointing away from us. The Doppler factor ($\delta$), Lorentz factor ($\Gamma$), intrinsic velocity ($\beta$) and jet orientation angle ($\theta$) are related by the following equation

\begin{equation}
\delta= \frac{1}{\Gamma (1-\beta cos{\theta})}.
\end{equation}
\label{Dopplerfactor}

With $\Gamma=1.5$, $\beta=0.76$ for the jet with the N components, the Doppler factor is 2.7 as given in \cite{Giroletti_2005} and for the jet with the S components it is 0.38.
\par
\begin{figure}[ht]
    \centering
    \includegraphics[width=0.9\linewidth]{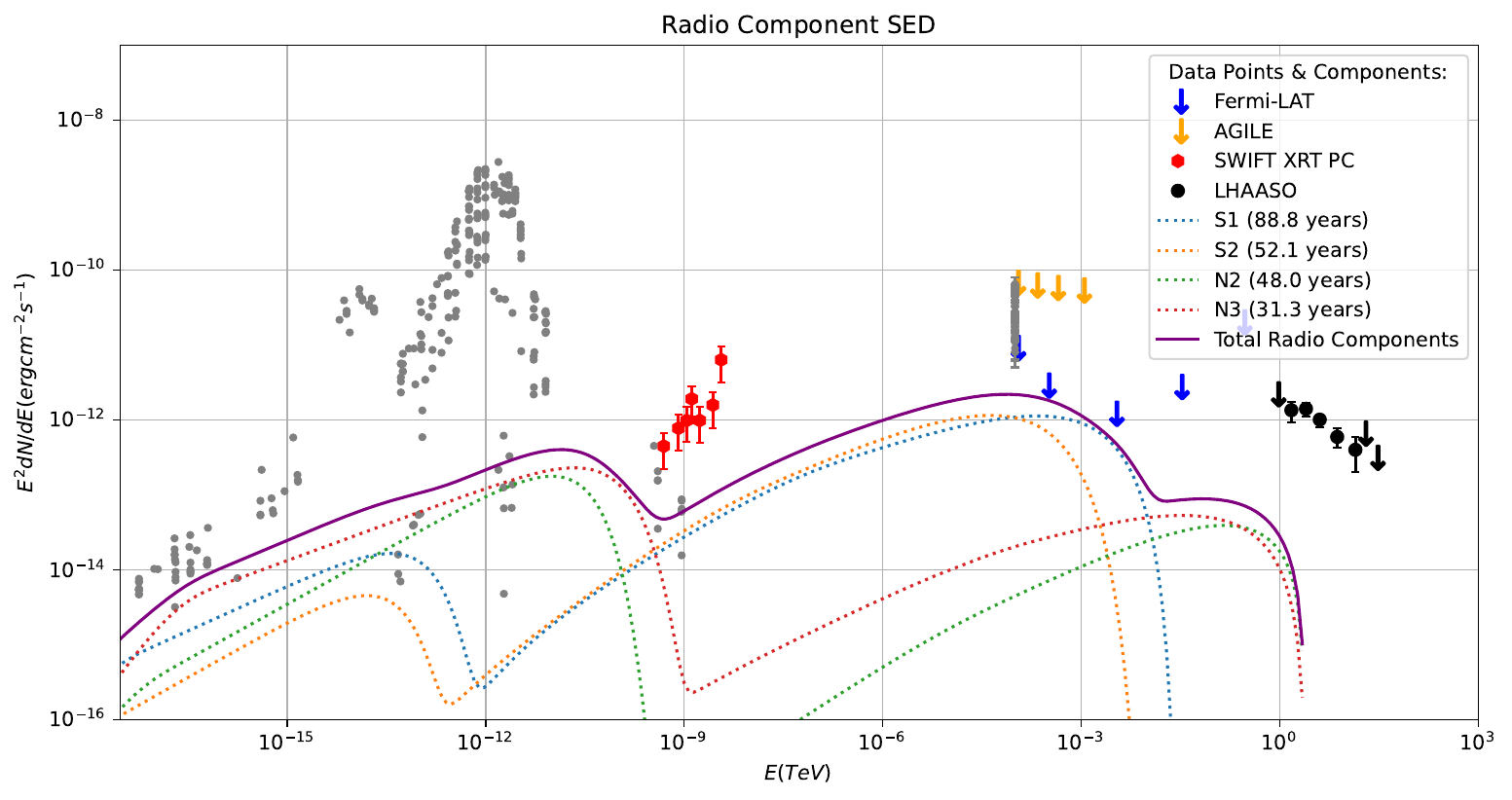}  
    \caption{SSC model used to obtain the multi-wavelength SED from each radio component in the year 2023 shown with dotted lines, their sum is shown with the solid line and compared with the observational data points as mentioned in Fig \ref{fig:time_n3}. 
    The values of the model parameters of the components used to obtain the SEDs are mentioned in Table \ref{tab:parameter}.}
    \label{fig:add_sed}
\end{figure}

We have assumed that the electrons are accelerated by the first-order Fermi mechanism hence their spectral index is close to -2. The normalisation of the electron spectrum is related to the jet power in electrons. The radio data points constrain the magnetic field and the normalisation. The values of the parameters are presented in Table \ref{tab:parameter}. In this way, we estimate the electron spectrum and magnetic field inside each component and subsequently evolve the electron spectrum to the current age of each component, to get the SED at the present epoch. The jet power required in injected electrons is of the order of $10^{40}$ erg/sec in N2 and N3, and $10^{43}$ erg/sec in S1 and S2 respectively and the magnetic field inside the radio components is lower than 1 mG. If we increase the maximum energies of the electrons in the N2 and N3 components then we cannot fit the Swift X-ray data points after adding the SEDs of all the components, and if we increase the maximum energies of the electrons in the S1 and S2 components then their SSC emissions become too high and the total SED exceeds the Fermi-LAT upper limits. Thus the maximum energies of the electrons in the radio components are constrained by the X-ray data points and the Fermi-LAT upper limits. The radio data points' energies constrain the electrons' minimum energies in the radio components. Fig \ref{fig:add_sed} shows the sum of the SEDs of the radio components at the present epoch, which cannot explain the Swift X-ray (\cite{2024ApJS..271...10W}) and LHAASO gamma-ray data (\cite{LHAASO:2024qzv}). The SSC emissions from S1 and S2 components may contribute to the gamma-ray spectrum near 100 MeV. In future with the detection of gamma rays by the Fermi-LAT detector, it would be possible to constrain the maximum energies of the electrons in the S1 and S2 components. Moreover, the double hump structure in the SED of NGC 4278 disappears if the S1 and S2 components emit gamma rays near 100 MeV.
\par
We have proposed that an additional component is required to explain the high energy emission data obtained from Swift-XRT PC \citep{2024ApJS..271...10W} and LHAASO \citep{LHAASO:2024qzv}. Parameters such as the normalisation of the injected electron spectrum, and magnetic field inside the component have been varied to fit the high-energy data. The spectral index is chosen to be -2 as before. The minimum and maximum energy of the injected electrons are assumed to be higher than those of the radio components to produce very high-energy secondary photons. The age of the component is assumed to be 30 years, which is comparable to the age of the youngest radio component N3. The Doppler factor of the comoving frame is 2.7, which is provided by \cite{Giroletti_2005}. The size of the high energy component is assumed to be much smaller (0.005 pc) compared to the size of the radio components so the SSC component is significantly higher and reaches the flux measured by LHAASO.  

\begin{figure}[ht]
    \centering
    \includegraphics[width=0.9\linewidth]{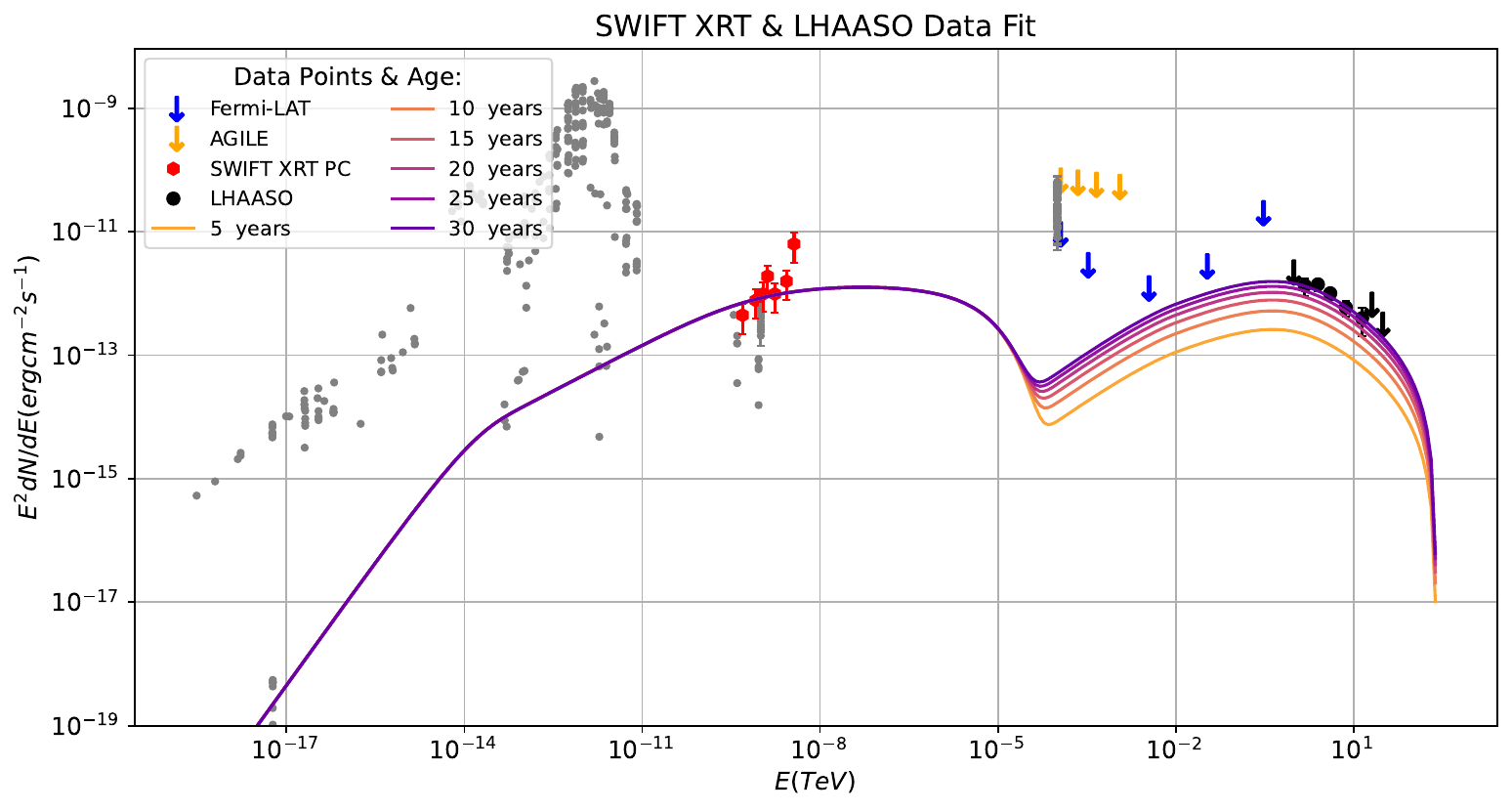}  
    \caption{
    High energy emission from a component in jet of NGC 4278 in the year 2023 assuming its age is 30 years which can explain the Swift X-ray \citep{2024ApJS..271...10W} and LHAASO gamma-ray data \citep{LHAASO:2024qzv}. The SSC model is used to get the multi-wavelength SED. The values of the model parameters used to get this SED are mentioned in Table \ref{tab:parameterhigh}. The time evolution of the SED is also shown.}
    \label{fig:he_power}
\end{figure}

\begin{figure}[ht]
    \centering
    \includegraphics[width=0.9\linewidth]{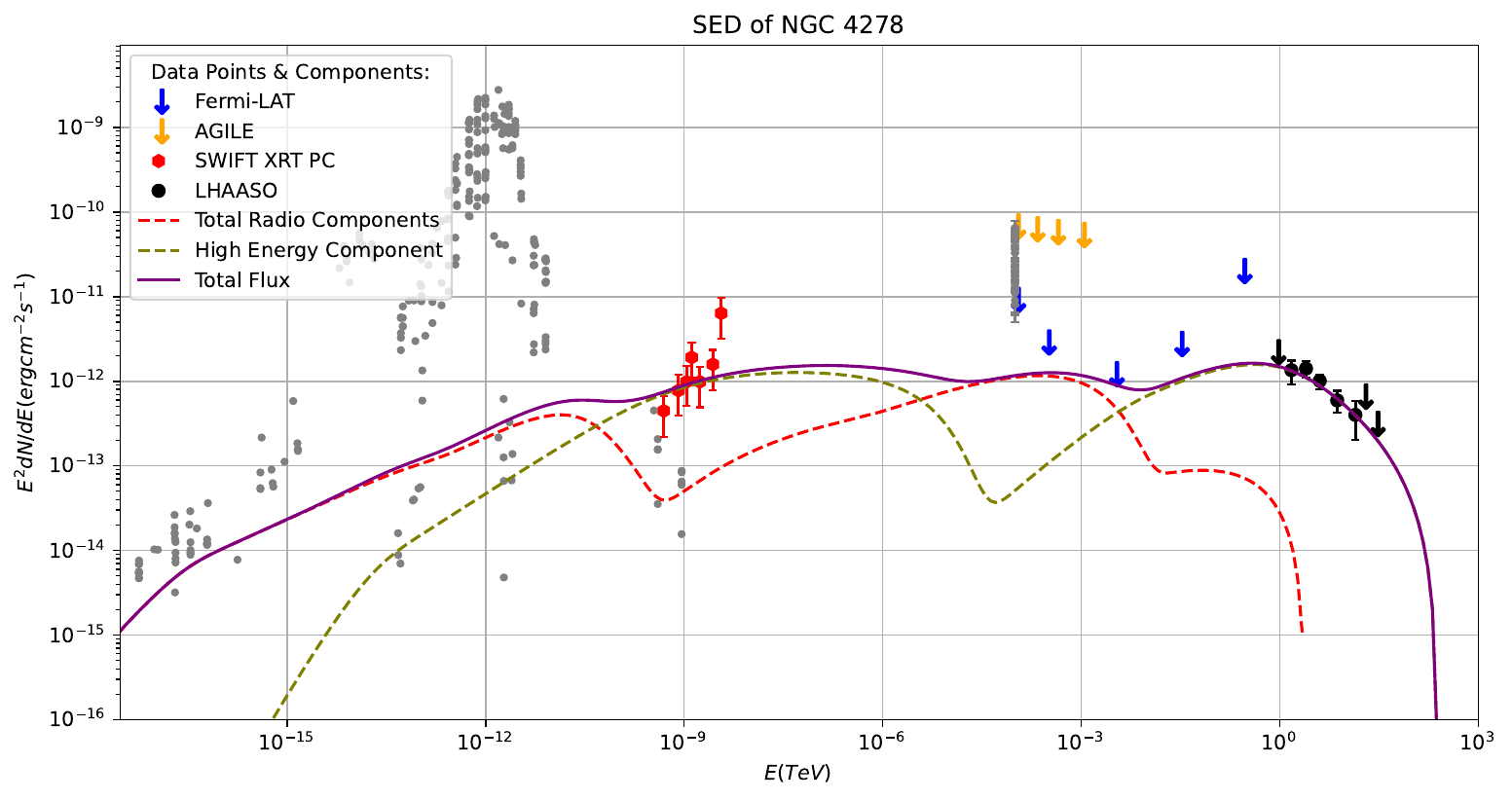}  
    \caption{The total SED in the year 2023 is calculated after adding the SEDs of the radio components shown in Fig \ref{fig:add_sed} to the SED of the high energy emission shown in Fig \ref{fig:he_power}, the comparison of our model prediction with multi-wavelength observational data is shown.}
    \label{fig:total_sed_power}
\end{figure}

The values of the various parameters used in our modelling have been mentioned in Table \ref{tab:parameterhigh}. We have evolved the electron spectrum to obtain the SED at the present epoch as shown in Fig. \ref{fig:he_power}. The jet power in injected electrons has to be $5 \times 10^{39}$ erg/sec, and the magnetic field inside the component is $8$ mG, these two parameters are constrained by the synchrotron and SSC emission which explains the Swift XRT and LHAASO data. The SED fits the Swift X-Ray data \citep{2024ApJS..271...10W}, it is below the Fermi-LAT upper limits, it also fits the LHAASO - WCDA measurements \citep{LHAASO:2024qzv}. The total SED after adding all the components is shown in Fig. \ref{fig:total_sed_power}.


\begin{deluxetable*}{ccccc}
\tablenum{1}
\tablecaption{Parameters used to fit the radio data of jet components of NGC 4278 with simple power law form of injected electron spectrum $Q(E/E_0)\propto \left(\frac{E}{E_0}\right)^{-\alpha}$.
\label{tab:parameter}}
\tablewidth{0pt}
\tablehead{
\colhead{Parameters} & \colhead{S2} &\colhead{S1} &\colhead{N3} &\colhead{N2}
}
\startdata
Power injected in electrons (in erg/sec) &$ 7.5 \times 10^{42}$ & $1.0 \times 10^{43}$ & 2.0 $\times 10^{40}$ & 2.25 $\times 10^{40}$ \\
Reference energy ($E_0$ in GeV)  & $ 10^{3}$ & $ 10^{3}$ & $10^{3}$ & $10^{3}$ \\
Maximum energy of electrons injected ($E_{max}$ in GeV) & 50 &  100 & $10^{3}$ & $10^{3}$ \\
Minimum energy of electrons injected ($E_{min}$ in GeV) &$ 0.16$  & $0.32$  & $ 0.32$  & $0.32$ \\
Spectral index of injected spectrum ($\alpha) $ & $ 2.0 $ &$2.17$ & $2.35$ & $2.02$\\
Magnetic field in emission region (B in mG) & $ 0.8 $ & $0.55$  & $0.9$ & $0.2$  \\
Size of emission region ($R$ in pc)\tablenotemark{a} & $0.0748$ & $0.3695$  & $0.18734$  & $0.5398$ \\
Escape time ($t_{esc}$ in sec) & $10R/c$  & $10R/c$  & $10R/c$  & $10R/c$  \\
Age of component (in 2000 in years)\tablenotemark{a} & $29.1 \pm 9.3$  & $65.8 \pm 12.4$  & $8.3 \pm 0.5$  & $25.0 \pm 4.8$ \\
Age of component (in 2023 in years) & $ 52.1$  & $88.8$ & $31.3$  & $48.0$ \\
Doppler factor of component ($\delta$)\tablenotemark{a} & $0.38$ & $0.38$ & $2.7$ & $2.7$ \\
Distance to the component (in Mpc)\tablenotemark{b} & 16.4 & 16.4 & 16.4 & 16.4 \\
\enddata
\tablenotetext{a}{\cite{Giroletti_2005}}\tablenotetext{b}{\cite{Tonry_2001}}
\end{deluxetable*}

\begin{deluxetable*}{cc}
\tablenum{2}
\tablecaption{Parameters used to fit the high energy data of NGC 4278 with power law form of injected electron spectrum $Q(E/E_0) \propto \left(\frac{E}{E_0}\right)^{-\alpha}$.
\label{tab:parameterhigh}}
\tablewidth{0pt}
\tablehead{\colhead{Parameters} & \colhead{High energy component}}

\startdata
Power injected in electrons (in erg/sec) & $5 \times 10^{39}$  \\
Reference energy($E_0$ in GeV)  & $10^{3}$ \\
Maximum energy of electrons injected ($E_{max}$ in GeV) & $10^{5}$\\
Minimum energy of electrons injected ($E_{min}$ in GeV) & 3.16 \\
Spectral index of injected spectrum ($\alpha$) &$2.0$\\
Magnetic field in emission region(B in mG) & $8.0$ \\
Size of emission region ($R$ in pc) & $0.005$\\
Escape time ($t_{esc}$ in sec) & $10R/c$ \\
Age of component (in 2023 in years) & $30.0$  \\
Doppler factor of component ($\delta$) & $2.7$ \\
Distance to the component (in Mpc)\tablenotemark{a} & $ 16.4$ \\
\enddata
\tablenotetext{a}{\cite{Tonry_2001}}
\end{deluxetable*}
\section{Discussion}

LLAGNs are much less explored compared to the high luminosity ones. The time evolution of different emission components in the jets may play an important role in understanding this class of objects. The association of radio and X-ray flares with the knot ejection from M 81* \citep{2016NatPh..12..772K} indicated a connection in the emission mechanisms of radio and X-ray photons. They are both produced in synchrotron emission of ultra-relativistic electrons \citep{2023ApJ...950..113T}. In the case of NGC 4278 also, the radio and X-ray emission may be from synchrotron emission of ultra-relativistic electrons. However, in this case, we do not have simultaneous flare data in radio and X-ray energy bands to show that the same population of electrons can explain the observed radio and X-ray data. Moreover, we see that the long-term data covering different frequency ranges cannot be explained with the same population of electrons, a separate emission region is required to explain the X-ray and gamma-ray observations. The gamma rays are produced in SSC emission in the model discussed in this work, which constrains the size of the emission region. In the case of M 81* there is no gamma-ray data, only upper limits are available at present, which can only put a lower bound on the size of the emission region. The very high energy gamma-ray data recorded from the direction of NGC 4278 has been used here to constrain the size of the emission region, which has to be of the order of $10^{16}$ cm to make the SSC bump very high to reach the gamma-ray flux measured by LHAASO - WCDA.

\par
The multi-wavelength data from NGC 4278 has been modelled earlier with leptonic and hadronic interactions by \cite{2024ApJS..271...10W}. Their one-zone SSC model uses a steady state, broken power-law electron density distribution as indicated in \cite{Xue_2019}. Their one-zone SSC model can fit the X-ray and very-high-energy gamma-ray data with extreme values of parameters. With an SSC + $p p$ model considering a steady state proton energy density, their SED fits all the data without requiring any extreme values of parameters. The sizes of the emission region in their SSC models are $8.5\times 10^{13}$ cm and $1.5\times 10^{14}$ cm. The sizes of the emission region in their SSC+$p p$ models are $10^{15}$ cm and $3\times 10^{15}$ cm. The internal optical depth of very high-energy gamma rays depends on the density of the low-energy photons, which absorb the very high-energy gamma rays. They have shown in their figures that the absorption of very high-energy gamma rays by the low-energy photons inside the jet is not important for NGC 4278. In our modelling, the size of the high energy emission region ($1.5\times 10^{16}$ cm) is larger than theirs, hence internal optical depth does not attenuate the very high gamma-ray flux significantly. Moreover, due to the proximity of NGC 4278, its very high energy gamma-ray flux is not attenuated by the extra-galactic background light. In our model, the magnetic field is lower than \cite{2024ApJS..271...10W} but the minimum and maximum energy of electrons are much higher at injection inside the high-energy component. These electrons have lost energy over 30 years, their time-evolved spectrum has been used to calculate the radiation spectrum. In their work, the escape timescale of electrons is $R/c$, which is 10 times lower than the value used in our work. Since the electrons are escaping faster and they have used the steady-state electron spectrum to fit the high energy observations, the luminosity in injected electrons is higher in their paper. The value of the Doppler factor used in \cite{2024ApJS..271...10W} for the small viewing angle is similar to the value used in our work. 

\par
The emission from the host galaxy of NGC 4278 is shown by the archival optical data points in our figures, which is also present in other LLAGNs e.g. NGC 315, NGC 1275 \citep{2021ApJ...919..137T} and M 81* \citep{2023ApJ...950..113T}. Multi-wavelength variability has been observed in these sources. The variability of M 81* in radio and X-ray frequencies has been modelled in \cite{2023ApJ...950..113T}. NGC 4278 also has variability in different frequencies (\cite{Younes_2010}, \cite{Lian:2024xnb}), however, a variability study is not the objective of this work. The latest LHAASO data points from \cite{LHAASO:2024qzv} have been included in our figures to compare with the simulated spectrum. The light curve of NGC 4278 shows a variability time-scale of a few months in TeV energy. A month-scale variability implies that the region size has to be less than $R\leq 2 \times 10^{17}$ cm, which is consistent with the size of the emission region $1.5\times 10^{16}$ cm used in our work. The magnetic field of the high energy emission region is estimated to be more than 5 mG in \citep{LHAASO:2024qzv}, which is consistent with the value (8 mG) used in our work. 
\par
Our study does not include the time evolution of the parameters of the components. The sizes of the components given in Table 2 of \cite{Giroletti_2005} do not scale with their distances from the centre of the S-shaped twin jets, hence it is not possible to use a conical geometry for the time evolution of the components. If the high-energy component is much younger than the youngest radio component N3 then a smaller size would be required to enhance the SSC emission to explain the LHAASO data points. If the high-energy component is older than 30 years then the high-energy SED does not change significantly with time, hence for the same size of the high-energy component, it would be possible to fit the gamma-ray data. The high-energy component has electrons of much higher energy compared to the radio components, otherwise, the very high-energy gamma-rays cannot be produced in the jet of NGC 4278. It is unknown what may cause the high-energy component to be different from the radio components. If the acceleration of electrons is more efficient in the vicinity of the high-energy component then there could be an injection of much more energetic electrons which may produce the high energy radiation by energy dissipation. Diffusive shock acceleration may become more efficient in turbulent reconnection, which is produced by large amplitude magnetic disturbances in shocks \citep{garrel_2018}.  In some cases, multiple zones or components are required to explain multi-wavelength data of blazars \citep{Liu_2023} in both low and high states. Sub-structures and bright knots have been observed along the jets of several blazars. VLBA image of M 87 jet shows many sub-structures having different proper motions \citep{mertens_2016}. Simultaneous radio, optical, X-ray and gamma-ray observations of 3C 279 revealed multiple energy dissipation regions in its jet \citep{rani_2018}. In several blazars, the flux variability in different energy bands cannot be correlated \citep{magic_2021}, and in some cases, orphan gamma-ray flares have been observed (\cite{Liodakis_2019}, \cite{magic_2019}, \cite{2021JHEAp..29...31P}), these results suggest there are multiple emission regions. Hence, different components in the jets of NGC 4278 indicate that LLAGNs may have multiple energy dissipation regions similar to the high luminosity AGNs.

\subsection{Comparison of NGC 4278 with Highly Peaked BL Lac objects}
\begin{figure}[ht]
    \centering
    \includegraphics[width=0.9\linewidth]{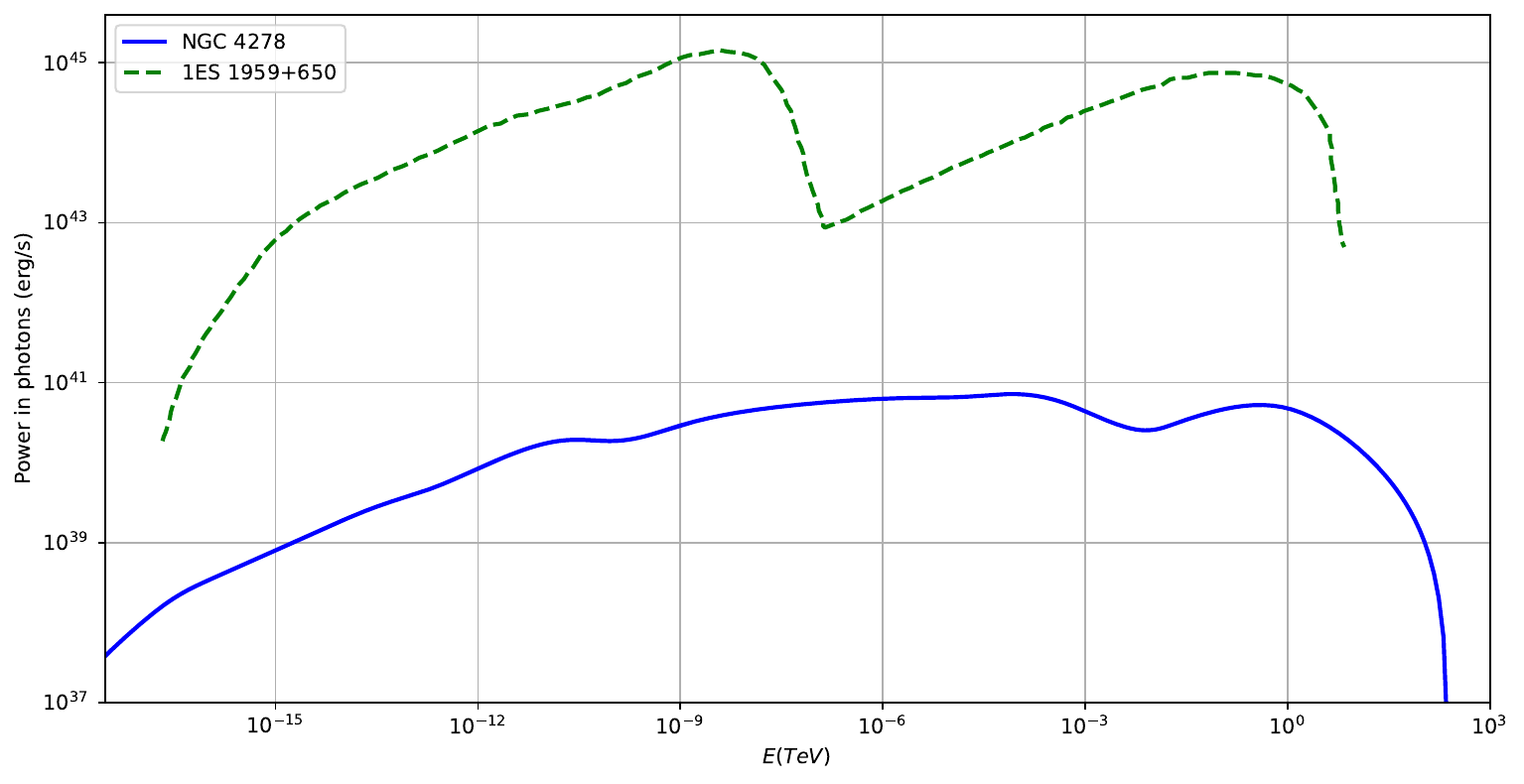}
    \caption{A comparison of the powers emitted in photons} of NGC 4278 (from this paper) and 1ES 1959+650 during its intermediate state \citep{acciari_2020}.
    \label{fig:compare_sed}
\end{figure}

\cite{acciari_2020} fitted the broadband spectra of five very high-energy BL Lac objects with one-zone and two-zone models. MAGIC gamma-ray telescope detected TeV gamma-rays from these sources, also variability of these sources was detected in different energy bands which helped the authors to model these sources. The one-zone model cannot fit the radio data though it can fit the optical to very high energy gamma-ray data. A larger emission region is required to fit the radio data compared to the data at higher frequencies. For NGC 4278 also the radio components are larger compared to the high-energy component and multiple components are required to fit the observational data covering the radio to very high-energy gamma-ray frequencies. We compare the SED of NGC 4278 to the SED of high energy peaked BL Lac object 1ES 1959+650 during its intermediate state in Fig. \ref{fig:compare_sed} as both emit very high energy gamma rays. The double hump structure cannot be seen in the SED of NGC 4278 if the S1 and S2 components emit significantly near 100 MeV energy. The Doppler factors of NGC 4278 and 1ES 1959+650 differ by a factor of approximately 10 and the powers emitted in photons near 1 TeV gamma-ray energy differ by approximately a factor of $ 10^4$. Higher variability is expected in high-energy BL Lac objects due to the higher values of the Doppler factors. We notice that NGC 4278 resembles the high-energy peaked BL Lac objects to some extent although it is less luminous and has a slow jet.
\section{Conclusion}
We conclude that our model fits the archival radio data, fits the X-ray data of Swift analysed by \cite{2024ApJS..271...10W} and the very high energy gamma ray data of LHAASO \citep{Cao_2024a} after adding the photon fluxes from multiple components as shown in Fig. \ref{fig:total_sed_power}. We have shown that the radio components S1, S2, N3 and N2 cannot explain the Swift XRT observations in the X-ray energy band and the high energy gamma ray data recorded by the LHAASO detector with their time-evolved electron spectrum. The S1 and S2 components may contribute to the gamma-ray spectrum near 100 MeV, which could be verified in future with Fermi-LAT data.
A separate component is needed in the jet to model the Swift XRT and LHAASO data points. Simultaneous radio, X-ray and gamma-ray observations are needed to study the individual components in the jets of NGC 4278 and other LLAGNs. Since they are closer to us it would be easier to monitor the individual jet components over many years to understand the underlying mechanisms of jet emission. We note that NGC 4278 has the following characteristics similar to some of the high luminosity AGNs. It has multiple emission regions and the sizes of the radio components are larger than the size of the high-energy component.

\section{Acknowledgment}
The authors thank the referee for helpful comments. They also thank Sovan Boxi for technical support related to software issues, Gunjan Tomar for helpful discussion and K. S. Dwarakanath for explaining the error analysis of radio flux measurement. This research has made use of the NASA/IPAC Extragalactic Database (NED), which is operated by the Jet Propulsion Laboratory, California Institute of Technology, under contract with the National Aeronautics and Space Administration. SD thanks BITS, Pilani for local hospitality where a part of this work was carried out.

\software{GAMERA (\url{http://libgamera.github.io/GAMERA/docs/documentation.html})}
\bibliographystyle{aasjournal}
\bibliography{Reference}

\end{document}